# A closer look at the hysteresis loop for ferromagnets
## - A survey of misconceptions and misinterpretations in textbooks

Hilda W. F. Sung and Czeslaw Rudowicz

Department of Physics and Materials Science, City University of Hong Kong, 83 Tat Chee Avenue, Kowloon, Hong Kong SAR, People's Republic of China

This article describes various misconceptions and misinterpretations concerning presentation of the hysteresis loop for ferromagnets occurring in undergraduate textbooks. These problems originate from our teaching a solid state / condensed matter physics (SSP/CMP) course. A closer look at the definition of the *'coercivity'* reveals two distinct notions referred to the hysteresis loop: B vs H or M vs H, which can be easily confused and, in fact, are confused in several textbooks. The properties of the M vs H type hysteresis loop are often ascribed to the B vs H type loops, giving rise to various misconceptions. An extensive survey of textbooks at first in the SSP/CMP area and later extended into the areas of general physics, materials science and magnetism / electromagnetism has been carried out. Relevant encyclopedias and physics dictionaries have also been consulted. The survey has revealed various other substantial misconceptions and/or misinterpretations than those originally identified in the SSP/CMP area. The results are presented here to help clarifying the misconceptions and misinterpretations in question. The physics education aspects arising from the textbook survey are also discussed. Additionally, analysis of the CMP examination results concerning questions pertinent for the hysteresis loop is provided.

## 1. Introduction

During years of teaching the solid state physics (SSP), which more recently become the condensed matter physics (CMP) course, one of us (CZR), prompted by questions from curious students (among others, HWFS), has realized that textbooks contain often not only common misprints but sometimes more serious misconceptions. The latter occur mostly when the authors attempt to present a more advanced topic in a simpler way using schematic diagrams. One such case concerns presentation of the magnetic hysteresis loop for ferromagnetic materials. Having identified some misconceptions existing in several textbooks currently being used for our SSP/CMP course at CityU, we have embarked on an extensive literature survey. Search of physics education journals have revealed only a few articles dealing with magnetism, e.g. Hickey & Schibeci (1999), Hoon & Tanner (1985). Interestingly, a review of middle school physical science texts by Hubisz (http://www.psrc-online.org/curriculum/book.html), which has recently come to our attention, provides ample examples of various errors and misconceptions together with pertinent critical comments. However, none of these sources have provided clarifications of the problems in question. To find out the extent of these misconceptions existing in other physics areas, we have surveyed a large number of available textbooks pertinent for solid state / condensed matter, general physics, materials science, and magnetism / electromagnetism. Several pertinent encyclopedias and physics dictionaries have also been consulted. The survey has given us more than we bargained for, namely, it has revealed various other substantial misconceptions than those originally identified in the SSP/CMP area. The results of this survey are presented here for the benefit of physics teachers (as well as researchers) and students. The textbooks, in which no relevant misconceptions and/or confusions were identified, are not quoted in text, however, they are listed for completeness in Appendix I in order to provide a comprehensive information on the scope of our survey.

In order to provide the counterexamples for the misconceptions identified in the textbooks, we have reviewed a sample of recent scientific journals searching for real examples of the magnetic hysteresis loop, beyond the schematic diagrams found in most textbooks. To our surprise a number of general misconceptions concerning magnetism have been identified in this review. The results of this review are presented in a separate article, which focuses on the research aspects and provides recent literature data on soft and hard magnetic materials (Sung & Rudowicz, 2002; hereafter referred to as S & R, 2002).

The root of the problem appears to be the existence of two ways of presenting the hysteresis loop for ferromagnets: (i) B vs H curve or (ii) M vs H curve. In both cases, the *'coercivity' ('coercive force')* is defined as the point on the negative H-axis, often using an identical symbol, most commonly $H_c$. Yet it turns out that the two meanings of *'coercivity'* are not equivalent. In some textbooks the second notion of coercivity (M vs H) is distinguished from the first one (B vs H) as the *'intrinsic'* coercivity $H_{ci}$. An apparent identification of the two meanings of coercivity $H_c$ (B vs H) and $H_{ci}$ (M vs H) as well as of the properties of soft and hard magnetic materials have lead to misinterpretation of $H_c$ as the point on the B vs H hysteresis loop where the magnetization is zero. This is evident, for example, in the statements referring to $H_c$ as the point at which *"the sample is again unmagnetized"* (Serway, 1990) or *"the field required to demagnetize the sample"* (Rogalski & Palmer, 2000). Other misconceptions identified in our textbook survey concern: 'saturation induction $B_{sat}$' and the inclination of the B vs H curve after saturation, shape of the hysteresis loop for soft magnetic materials, and presentation of the hysteresis loop for both soft and hard ferromagnets in the same diagram. Minor problems concerning terminology and the drawbacks of using schematic diagrams are also discussed. Analysis of the condensed matter physics examination results concerning questions pertinent for the hysteresis loop is provided to illustrate some popular misconceptions in students' understanding.

## 2. Two notions of coercivity

For a ferromagnetic material, the magnetic induction (or the magnetic field intensity) inside the sample, **B**, is defined as (see, e.g. any of the books listed in References):

$$\mathbf{B} = \mathbf{H} + 4\pi \mathbf{M} \quad \text{(CGS)};$$
$$\mathbf{B} = \mu_o ( \mathbf{H} + \mathbf{M} ) \quad \text{(SI)} \quad (1)$$

where **M** is the magnetization induced inside the sample by the applied magnetic field **H**. In the free space: $\mathbf{M} = 0$ and then in the SI units: $\mathbf{B} \equiv \mu_o \mathbf{H}$, where $\mu_o$ is the permeability of free space ($\mu_o = 4\pi \times 10^{-7}$ [m kg A$^{-2}$ sec$^{-2}$]; note that the units [Hm$^{-1}$] and [WbA$^{-1}$m$^{-1}$] are also in use). The standard SI units are: B [tesla] = [T], H and M [A/m], whereas B [Gauss] = [G], H [Oersted] = [Oe], and M [emu/cc] (see, e.g. Jiles, 1991; Anderson, 1989). Both the CGS units and the SI units are provided since the CGS unit system is in use in some textbooks surveyed and comparisons of values need to be made later.

In Fig. 1 we present schematically the hysteresis curves for a ferromagnetic material together with the definitions of the terms important for technological applications of magnetic materials. The two meanings of *'coercivity'* $H_{ci}$ and $H_c$ as defined on the diagrams: (a) the magnetization M vs applied field H and (b) magnetic induction (or flux density) B vs H, respectively, are clearly distinguished. Both curves have a similar general characteristic, except for one crucial point. After the saturation point is reached, the M curve becomes a straight line with exactly zero slope, whereas the slope of the B curve reflects the constant magnetic susceptibility and depends on the scale and units used to plot B vs H (see below). In other words, the B vs H curve does not saturate by approaching a limiting value as in the case of the M vs H curve.

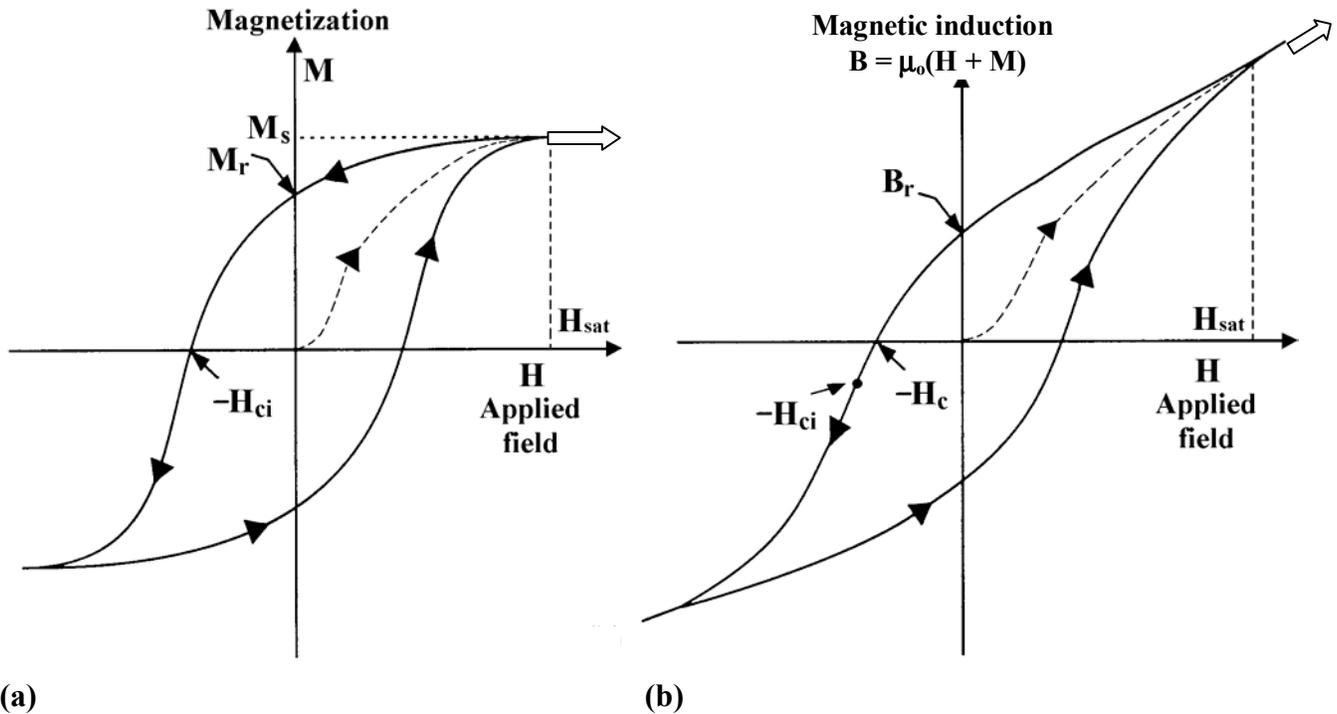

(a)                                                                 (b)

**Fig. 1.** Hysteresis curves for a ferromagnetic material: (a) **M vs H**: $M_r$ is the remanent magnetization at H = 0; $H_{ci}$ is the *intrinsic* coercivity, i.e. the reverse field that reduces M to zero; $M_s$ is the saturation magnetization; (b) **B vs H**: $B_r$ is the remanent induction (or *'remanence'*) at H = 0; $H_c$ is the coercivity, i.e. the reverse field required to reduce B to zero (adapted from Elliot, 1998).

For an initially unmagnetized sample, i.e. M ≡ 0 at H = 0, as H increases from zero, M and B increases as shown by the dashed curves in Fig. 1 (a) and (b), respectively. This magnetization process is due to the motion and growth of the **magnetic domains**, i.e. the areas with the same direction of the local magnetization. For a full discussion of the formation of hysteresis loop and the nature of magnetic domains inside a ferromagnetic sample one may refer to the specialized textbooks listed in the References, e.g. Kittel (1996), Elliott (1998), Dalven (1990), Skomski & Coey (1999). Here we provide only a brief description of these aspects. A distinction must be made at this point between the magnetically **isotropic** materials, for which the magnetization process does not depend on the orientation of the sample in the applied field **H**, and the **anisotropic** ones, which are magnetized first in the **easy** direction at the lower values of **H**. In the former case, as each domain magnetization tends to rotate to the direction of the applied field Kittel (1996), the domain wall displacements occur, resulting in the growth of the volume of domains favorably oriented (i.e. parallel) to the applied field and the decrease of the unfavorably oriented domains Kittel (1996). In the latter case, only after the **magnetic anisotropy** (for definition, see, e.g. Kittel (1996), Elliott (1998), Dalven (1990), Skomski & Coey (1999), Jiles (1991)) is overcome the sample is fully magnetized with the direction of **M** along **H**. In either case, when this *'saturation point'* is reached, the magnetization curve no longer retraces the original dashed curve when H is reduced. This is due to the irreversibility of the domain wall displacements. When the applied field H reaches again zero, the sample still retains some magnetization due to the existence of domains still aligned in the original direction of the applied field Dalven (1990). The respective values at H = 0 are defined (see, e.g. Kittel (1996), Elliott (1998), Dalven (1990), Skomski & Coey (1999), Jiles (1991)) as the **remnant magnetization** $M_r$, Fig. 1 (a), and the **remnant induction** $B_r$, Fig. 1 (b). To reduce the magnetization M and magnetic induction B to zero, a reverse field is required known as the **coercive force** or **coercivity**. The **soft** and **hard** magnetic materials are distinguished by their small and large area of the hysteresis loop, respectively.

By definition, the coercive force (coercivity) defined in Fig. 1 (a), and that in Fig. 1 (b) are two different notions, although their values may be very close for some materials. In order to distinguish them, some authors define either the **related coercivity** (Kittel, 1996) or the **intrinsic coercivity** (Elliott, 1998; Jiles, 1991) $H_{ci}$ as the reverse field required to reduce the magnetization M from the remnant magnetization $M_r$ again to zero as shown in Fig. 1 (a), whereas reserve the symbol $H_c$ and the name **coercivity** (coercive force) to denote the reverse field required to reduce the magnetic induction in the sample B to zero as shown in Fig. 1 (b), as done, e.g. by Kittel (1996). Hence, the confusion between the two notions of coercivity referred to the curve B vs H and the curve M vs H can be avoided. Since a clear distinction between $H_c$ and $H_{ci}$, is often not the case in a number of textbooks, a question arises under what conditions and for which magnetic systems, if any, $H_c$ and $H_{ci}$ can be considered as equivalent quantities. If it was the case, the point $-H_c$ on the B vs H curve would also correspond to the magnetization M ≡ 0 as in the case of $H_{ci}$ on the M vs H curve. Only in one of the books surveyed such approximation is explicitly considered. Dalven (1990) shows that, in general, the values of B and M are much larger than H in both curves in Fig. 1. Hence, if H can be neglected in Eq. 1, then $B \approx \mu_o M$. This turns to be valid only for low values of H and the narrow hysteresis loop pertinent for the soft magnetic materials. In other words, the value of $H_c$ and $H_{ci}$ are indeed **very close**, so not identical, *for the soft magnetic materials only*. In this case $H_{ci}$ in Fig. 1 (a) and $H_c$ in Fig. 1 (b) can be considered as two equivalent points and hence $M \approx 0$ at $H_c$ as well.

The real examples of the magnetic hysteresis loop, identified in our review (S & R, 2002) of a sample of recent scientific journals, indicate that $H_c$ and $H_{ci}$ turn out to be significantly **non-equivalent** *for the hard magnetic materials*. In the article (S & R, 2002) we have also complied values of $H_{ci}$, $H_c$, and $B_r$ for several commercially available permanent magnetic materials revealed by our recent Internet search. These data indicate that although $H_c$ and $H_{ci}$ are of the same order of magnitude, in a number of cases $H_{ci}$ is substantially larger than $H_c$. Hence, in general, it is necessary to distinguish between $H_c$ and $H_{ci}$. Moreover, as a consequence of $H_{ci} \neq H_c$, the magnetization does not reach zero at the point $-H_c$ on the B vs H curve but at a larger value of $H_{ci}$ indicated schematically in Fig. 1 (b). However, in the early investigations of magnetic materials, before the present day very strong permanent magnets become available, the values of $H_c$ and $H_{ci}$ were in most cases not distinguishable. As the advances in the magnet technology progressed, more and more hard magnetic materials have been developed, for which the distinction between $H_{ci}$ and $H_c$ is quite pronounced (see Table 1 in S & R (2002)). The presentation in most textbooks reflects the time lag it takes for new materials or ideas to filter from scientific journals into the textbooks as 'schematically presented established knowledge'.

## 3. Results of textbooks survey

In our survey of the presentation of the hysteresis loop for ferromagnetic materials, in total about 300 textbooks in the area of solid state / condensed matter, general physics, materials science, magnetism / electromagnetism as well as several encyclopedias and physics dictionaries available in City University library were examined. We have identified around 130 books dealing with the hysteresis loop. In order to save the space an additional list of the books surveyed (37 items), which deal with the hysteresis loop in a correct way but are not quoted in the References, is available from the authors upon request.

It appears that from the points of view under investigation, generally, the encyclopedias and physics dictionaries contain no explicit misconceptions. This is mainly due to the fact that the hysteresis loop is usually presented at a rather low level of sophistication (see, e.g. Lapedes (1978), Lord (1986), Meyers (1990), Besançon (1985), Parker (1993)). However, in a few instances in the same source book both types of hysteresis loop (B vs H and M vs H) are discussed in separate articles written by different authors without clarifying the distinct notions, which may also lead to confusion. Examples include, e.g. (a) Anderson & Blotzer (1999) and Vermariën *et al* (1999), and (b) Arrott (1983), Donoho (1983), and Rhyne (1983). Hence, these authoritative sources could not help us to clarify the intricacies we have encountered. This have been achieved by consulting more advanced books on the topic, e.g., Kittel (1996), Dalven (1990), Skomski and Coey (1999), and/or regular scientific journals (for references, see, S & R, 2002).

Only a small number of books surveyed contain both types of the curves: B vs H and M vs H as well as provide clarification of the terminology concerning $H_c$ and $H_{ci}$ - Kittel (1996), Elliott (1998), Dalven (1990), Skomski & Coey (1999), Jiles (1991), Arrott (1983), Donoho (1983), Rhyne (1983), Levy (1968), Anderson & Blotzer (1999), Vermariën *et al*. (1999). Barger & Olsson (1987) provide both graphs but terminology is only referred to the B vs H graph. Most books deal only with one type of the hysteresis loop. The B vs H curve, which is more prone to misinterpretations, has been used more often in the surveyed books in all areas. A few books deal with the M vs H curve and provide, with a few exceptions (see Section C below), correct description and graphs (see, e.g. Lovell *et al*, 1981; Aharoni, 1996; Wert & Thomson, 1970; Elwell & Pointon, 1979). On the other hand, the M vs H curve is dominant in research papers surveyed (S & R, 2002). Surprisingly, while most of the textbooks surveyed attempt to adhere to the SI units, all but a few research articles reviewed still use the CGS units. This in itself is a worrying factor (S & R, 2002). The various misconceptions and/or misinterpretations identified in the course of our comprehensive survey of textbooks can be classified into five categories. Below we provide a systematic review of the books with respect to the problems in each category.

### A. Misinterpretation of the coercivity $H_c$ on the B vs H curve as the point at which M=0.

This was the original problem which has triggered the textbook survey. Various examples of this misinterpretation, consisting in ascribing 'zero magnetization' to the point $–H_c$ **on the B vs H** hysteresis loop, are listed below with the nature of the problem indicated by the pertinent sentences quoted.

**Solid state / condensed matter physics books**

- *"The magnetic field has to be reversed and raised to a value $H_c$ (called the coercive force) in order to push domain walls over the barriers so that we regain zero magnetization."* (Wilson, 1979)
- *"The point at which B=0 is the coercive field and is usually designated as $H_c$. It represents the magnetic field required to demagnetize the specimen."* (Pollock, 1990)
- *"The reverse field required to demagnetize the material is called the coercive force, $H_c$."* (Pollock, 1985)
- *"To remove all magnetization from a specimen then requires the application of a field in the opposite direction termed the coercive field."* (Elliott & Gibson, 1978)
- *"H at c is called the coercive force and is a measure of the field required to demagnetize the sample."* (Rogalski & Palmer, 2000)

**General physics books**

- *"The coercive force is a measure of the magnitude of the external field in the opposite*

- *direction needed to reduce the residual magnetization to zero."* (Ouseph, 1986)
- *"In order to demagnetize the rod completely, H must be reversed in direction and increased to $H_d$, the coercive force."* (Beiser, 1986)
- *"If the external field is reversed in direction and increased in strength by reversing the current, the domains reorient until the sample is again unmagnetized at point c, where B=0."* (Serway, 1990)
- *"...the magnetization does not return to zero, but remains (D) not far below its saturation value; and an appreciable reverse field has to be applied before it is much reduced again (E)." [where E corresponds to $H_c$ in Fig. 1 (b), and later]...."the field required to reverse the magnetization (point E on the graph) varies..."* (Akril *et al*, 1982)

**Materials science and magnetism / electromagnetism books**

- *"In order to destroy the magnetization, it is then necessary to apply a reversed field equal to the coercive force $H_c$."* (Anderson *et al*, 1990)
- *"To reduce the magnetisation, B, to zero the direction of the applied magnetic field must be reversed and its magnitude increased to a value $H_c$."* (John, 1983) Note here the symbol B is confusingly used for the magnetization as discussed later.
- *"If the H field is now reversed, the graph continues down to R in the saturated case. This represents the H field required to make the magnetization zero within a saturation loop and is termed the coercivity of the material."* (Compton, 1986)
- *"... the value of H when B=0 is called the coercivity, $H_c$; ... It follows that the coercivity $H_c$ is a measure of the field required to reduce M to zero."* (Dugdale, 1993)
- *"Note that an external field of strength $-H_c$, called the coercive field, is needed to obtain a microstructure with an equal volume fraction of domains aligned parallel and antiparallel to the external field (i.e., B = 0)."* (Schaffer *et al*, 1999)

Apparently, all the above quotes refer to the **intrinsic coercivity** $H_{ci}$ as defined on the M vs H curve, whereas the B vs H curve was, in fact, used to explain the properties of the hysteresis loop. Neither a proper explanation about the validity of the approximation $H_c \approx H_{ci}$ nor information on the type of ferromagnetic materials described by a given schematic hysteresis loop was provided in all the quotation cases. Hence, such statements constitute misconceptions, which could be avoided if the authors defined the term *'coercive force'* / *'coercivity'* as the reverse field required to demagnetize (M = 0) the ferromagnetic material sample with a reference to the M vs H curve. Otherwise, when referring to the B vs H curve, the quantity $H_c$ should rather be defined as the field required to bring the magnetic induction, instead of the magnetization, to zero. The description in the text and the curve used in the books cited above, simply imply that both B and M were equal to zero at the same value of H, i.e. $H_c$. However, since **B = $\mu_o$ ( H + M )**, when B = 0, M is equal to $-H_c$. Only when $H_c$ is very small, as it is the case for soft magnetic materials, the approximation M ≈ 0 at B =

0 and $H_c \approx H_{ci}$ holds. Without explicitly stating the necessary conditions for the validity of such approximation, the presentations of the hysteresis loop expressed in the above quotes convey an incorrect concept of the zero magnetization at the point -$H_c$ on the B vs H curve as applicable to any kind of ferromagnetic materials.

To predict the value of H on the B vs H curve for which in fact M = 0, we consider **M = B/$\mu_o$ – H**. In the second quadrant of the hysteresis loop (see Fig. 1), we have $-H_c \leq H \leq 0$, and hence M diminishes from M = $B_r/\mu_o$ at H = 0 to the nonzero value at -$H_c$, i.e. M = -$H_c$. This means that the direction of the magnetization is still opposite to that of the applied field. Further increase of the negative $H_c$ in the third quadrant on the B vs H curve yields M = 0 at H = –$H_{ci}$. This is why the value of $H_{ci}$ on the M vs H curve is always greater than that of $H_c$ on the B vs H curve. This relationship is indicated schematically by a dot (the point -$H_{ci}$) in Fig. 1 (b). The values in Table 1 in S & R (2002) illustrate that for strong permanent magnets $H_{ci}$ is substantially larger in magnitude than $H_c$.

### B. *Misconceptions concerning the meaning of the saturation induction $B_{sat}$*

Apart from the two notions of coercivity, the term of *'saturation induction'* is also prone to confusion. If this term is not defined properly, various misconceptions may arise. Usually, in most textbooks the term *'saturation'* refers to the *'process'* and thus the corresponding quantities exhibit no further change after a certain limit is reached. For instance, a sponge no longer absorbs any more water after full *'saturation'*. Similarly, the magnetization in ferromagnetic materials does not change after the saturation point is reached at $H_{sat}$ (see, Fig. 1). Since M becomes constant, M = $M_s$, further increase of the applied field H no longer changes the value of the magnetization M, as represented by the straight dotted line in Fig. 1(a). However, this is not the case for the induction B. According to Eq. (1) after the saturation point is reached at $H_{sat}$, B still increases with H. Confusion may occur if the term *'saturation'* is used with respect to the B vs H curve. In this case, the *'saturation induction'* $B_{sat}$ reflects that in the magnetization saturation process a certain limit has been reached, denoted by a particular point on the B vs H curve. But it **does not mean** that B has reached a definite limit like in the case of M. Correct descriptions are found in, e.g. Kittel (1996) who refers the *'saturation induction'* to the point on the B vs H graph at which the magnetization reaches a certain limit; Hammond (1986): *"as H is increased, B increases less and less. It reaches an almost constant saturation value"*. However, the value of B still increases if H is continuously applied to the sample after saturation of magnetization, no matter how small the value of $\mu_o H$ is as compared with $\mu_o M$. This distinction between the properties after saturation of the M vs H curve and those of the B vs H curve is often misrepresented as shown in Fig. 2 (see, e.g. Pollock (1990, 1985), Compton (1986), Flinn & Trojan (1990)). The shape of the B vs H curves apparently resembles closely the shape of the M vs H curve with a (nearly) straight horizontal line after saturation.

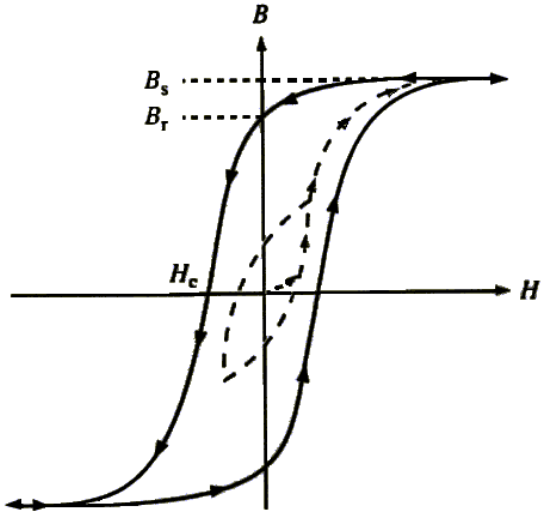

**Fig. 2.** Hysteresis loop for a ferromagnetic material with the saturation point indicated (adapted from Flinn & Trojan, 1990).

The misconception conveyed by such diagrams as in Fig. 2 is that after saturation, even if H increases further, the induction behaves in the same way as the magnetization, i.e. $B = B_s$ = const as $M = M_s$ = const. Such misconception is evident in a number of texts, for instance, *"With further increase in field strength, the magnitude of induction levels off at a saturation induction, $B_s$."* (Shackelford, 1996), *"This maximum value of B is the saturation flux density $B_s$"* (Callister, 1994), *"$B_{max}$ is the maximum magnetic induction…"* (Jastrzebski, 1987). The descriptions used in several other textbooks also reflect similar incorrect interpretation of the saturation induction $B_{sat}$, see, e.g. Pollock (1990, 1985), Compton (1986), Flinn & Trojan (1990), Van Vlack (1982), Selleck (1991), Harris & Hemmerling (1980), Arfken *et al* (1984), Knoepfel (2000), Brick *et al* (1977), and John (1983). Besides, in some surveyed books, the saturation is indicated incorrectly on the B vs H curve, e.g. Buckwalter *et al* (1987), Whelan and Hodgson (1982), Brown *et al* (1995). Those misconceptions could be avoided if a proper clarification is provided. It is then necessary to mention that, in fact, the contribution of the H term to B in Eq. (1) can be neglected but only for soft magnetic materials as they become saturated at small values of H. Some authors Ralls *et al* (1976), Burke (1986), Cullity (1972), Anderson *et al* (1990), Van Vlack (1970) have explicitly adapted this point of view. The term *'saturation induction'* is then used either under the assumption that after saturation of the magnetization H contributes to B in a negligible way, see, e.g. Ralls *et al* (1976), Anderson *et al* (1990), Van Vlack (1970), or it is not worth to increase B in the actual practice, see, e.g. Burke (1986). The former is true only for soft magnetic materials. However, the situation is quite different for hard magnetic materials for which, as it can be seen from Table 1 in S & R (2002), the values of coercivity $H_c$ (in kOe) are close to those of the remanence $B_r$ (in kGs), whereas the intrinsic coercivity $H_{ci}$ (in kOe) is greater than $B_r$ up to three times.

### C. Misconceptions concerning the actual inclination of the B vs H and/or M vs H curve after saturation

The misconceptions of this category, closely related to the category B, concern both the B vs H graphs and the M vs H graphs. Misconceptions may arise concerning the actual inclination if on the B vs H graphs the shape of the hysteresis loops resembles closely the shape of the M vs H loops and the B-lines after saturation appear to be represented by a straight horizontal line with zero inclination. If no proper explanation is provided the apparent zero inclination may be taken as a general feature of both graphs applicable to all magnetic materials. The opposite cases arise if on the M vs H graphs the shape of the hysteresis loops resembles closely the shape of the B vs H loops and the B-lines after saturation appear to be represented by lines with a **noticeable** inclination. Such cases amount to mixing up the M vs H graphs with the B vs H graphs and constitute misconceptions concerning the features of the M vs H hysteresis loops. Several cases of both versions of the misconceptions of this category have been revealed by considering the shape, inclination, and description of the B vs H and M vs H graphs in the textbooks.

The misconceptions concerning the B vs H graphs arise from the neglect of the difference between the *actual* and *apparent* inclination of the B-lines after saturation. After the magnetization saturation is reached, M becomes constant: $M = M_s$. Thus in the CGS units a further increase of H by 1 Oersted increases B by 1 Gauss, whereas in the SI units, correspondingly, 1 A/m of H contributes $4\pi \times 10^{-7}$ Tesla (i.e. the value of $\mu_o$) to B. Hence, no matter which units are used for the y- and x-axis, B is exactly proportional to H and must be represented by a straight line. However, the appearance of a graph depends on the actual inclination of the B-line after saturation, which is determined by the unit elements chosen for the y-axis ($y_{unit}$) and x-axis ($x_{unit}$), i.e. the scale used for the graph. To illustrate how the extension line of the B vs H hysteresis loop after saturation would look like for different scales, we have simulated the inclination corresponding to various scales used for the x- and y-axis as shown in Fig. 3. The lines S1 to S6 represent the unit element of the y-axis ($y_{unit}$) diminished by a factor of 1, 2, 5, 10, 100 and 1000 times, respectively. Thus the ratio $S = y_{unit} : x_{unit}$ (i.e. the re-scaling factor) equal to 1, 0.5, 0.2, 0.1, 0.01, and 0.001 yields the inclination 45°, 26.6°, 11.3°, 5.7°, 0.57°, and 0.057° for the lines S1, S2, S3, S4, S5, and S6, respectively. The same re-scaling factors apply if equivalently the unit element of the x-axis ($x_{unit}$) is increased. This is the case of the graphs where on the x-axis instead of H the quantity $B_o = \mu_o H$ is used. Then the units of Tesla are used on both the x- and y-axis, however, the typical values of $B_o$ are very small as shown in the second part of Table 1 below.

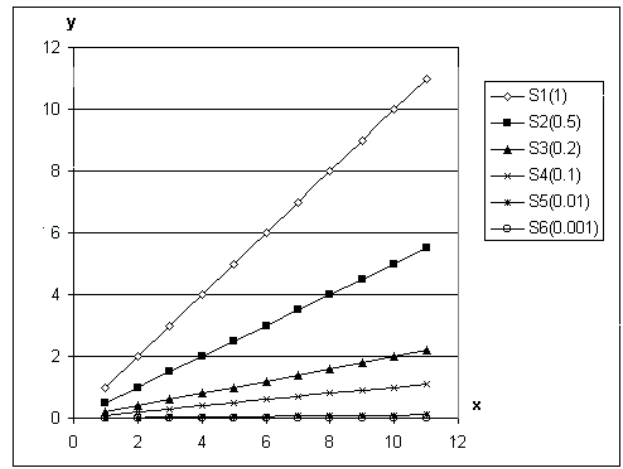

**Fig. 3.** Inclination of the B-line on the B (y) vs H (x) graph after magnetization saturation for various scales. The re-scaling factor: S1 (1), S2 (0.5), S3 (0.2) S4 (0.1), S5 (0.01), and S6 (0.001) corresponds to the inclination, i.e. the angle between the B-line and the x-axis, equal to: 45°, 26.6°, 11.3°, 5.7°, 0.57°, and 0.057°, respectively.

**Table 1.** Summary data extracted from the surveyed textbooks concerning the B vs H hysteresis graphs where the units were indicated. Data in brackets are in the CGS units; $X_{max}$ and $Y_{max}$ is the maximum value on the X- and Y-axis, respectively; the ratios $B_r$ (remanence) / $H_c$ (coercivity) or the equivalent ones are given as the order of magnitude only; the type of the inclination after saturation refers to the lines in Fig. 3; "—" means that no information was provided in the textbook.

| X-axis | | Y-axis | | $X_{max}$ | $Y_{max}$ | $B_r$ | $H_c$ | Ratio $B_r/\mu_0 H_c$ | Inclination type | Material name; type (soft / hard) | Reference |
|---|---|---|---|---|---|---|---|---|---|---|---|
| Name and / or symbol | SI units | Name and / or symbol | SI units | | | | | | | | |
| current H | A/m | magnetic induction B | T | 200 | 1 | 0.64 | 49 | $10^4$ | <S6 | — | Lorrain et al (1979) |
| magnetic intensity H | A/m | magnetic field B | T | 1600 | 1.6 | 1.25 | 1000 | $10^3$ | S6 | hard steel | Grant et al (1990) |
| magnetic intensity H | A/m | magnetic field B | T | 600 | 1.57 | 1.4 | 100 | $10^4$ | <S6 | commercial iron; hard | Grant et al (1990) |
| magnetic field intensity H | A/m | magnetic induction B | T | 1800 | 0.3 | 0.25 | 300 | $10^3$ | S2 | 4C4 ferrite at 25°C (soft) | Lea et al (1997) |
| magnetic field intensity H | A/m | magnetic induction B | T | $4 \times 10^5$ | 0.41 | 0.39 | $1.8 \times 10^5$ | $10^0$ | S1 | ferroxdur $BaFe_{12}O_{18}$; hard | Lea et al (1997) |
| magnetic field strength H | A/m | flux density B | T | 2000 | 0.5 | 0.3 | 600 | $10^2$ | S2 ~ S3 | cast iron; hard | Hammond (1986) |
| magnetic field strength H | A/m | flux density B | T | 2000 | 1.4 | 1 | 200 | $10^3$ | <S6 | mild steel; soft / hard | Hammond (1986) |
| magnetic field strength H | A/m | flux density B | T | 2000 | 1.4 | 0.8 | 20 | $10^4$ | <S6 | silicon iron; soft | Hammond (1986) |
| magnetic field strength H | A/m | magnetic induction B | Web/m² | $6 \times 10^4$ | 0.65 | 0.58 | $4 \times 10^4$ | $10^1$ | S4 | copper-nickel-iron alloy; hard | Shackelford (1996) |
| field intensity | A/m | flux density B | Web/m² | 750 | 1.8 | 1.3 | 120 | $10^3$ | <S6 | — | Harris et al (1980) |
| applied field H | A/m | flux density B | Web/m² | $48 \times 10^2$ | 1.5 | 0.6 | $20 \times 10^2$ | $10^2$ | S2 ~ S3 | steel; hard | Nelkon et al (1977) |
| applied field H | A/m | flux density B | Web/m² | $4 \times 10^2$ | 1.5 | 1.4 | $4 \times 10^2$ | $10^3$ | <S6 | soft iron; soft | Nelkon et al (1977) |
| magnetic intensity H | A/m | magnetic induction B | T | 160 | 1 | 0.5 | 28 | $10^4$ | <S6 | Silicon; soft | Granet (1980) |
| H | mT/$\mu_0$ | magnetic induction B | T | 25 | 1.9 | 1.1 | 6 | $10^2$ | S5 ~ S6 | — (hard) | Knoepfel (2000) |
| H | mT/$\mu_0$ | magnetic induction B | T | 0.2 | 1.4 | 0.2 | 0.2 | $10^3$ (*) | <S6 | — (soft) | Knoepfel (2000) |

Table 1. (cdn)

| | | | | $B_r$ ($B_r$) | $B_c$ | $B_r/B_c$ | | |
|---|---|---|---|---|---|---|---|---|
| $B_E$ | T | magnetic field B | 0.0006 | 1.4 | 1.25 | $12.5 \times 10^{-5}$ | $10^4$ | $< S6$ | commercial iron; soft | Arfken et al (1984) |
| $B_E$ | T | magnetic field B | 0.003 | 1.25 | 1.2 | 0.0015 | $10^2$ | $S5 \sim S6$ | tungsten steel; hard | Arfken et al (1984) |
| magnetising field $B_o$ external magnetic field $\mu_o H$ | T | B | 0.7 | 1.5 | 1.1 | $0.25 \times 10^{-3}$ | $10^3$ | $< S6$ | mild steel; hard | Akrill et al (1982) Fishbane et al (1993) |
| | T | internal magnetic field B | $3 \times 10^{-4}$ | 1.2 | 1.1 | $1.6 \times 10^{-4}$ | $10^3$ | $< S6$ | — | |
| external field $B_o$ | T | total magnetic field B | $1.2 \times 10^{-3}$ | 1 | 0.81 | 0.61 | $10^0$ | $S1 \sim S2$ | — | Giancoli (1991 & 1989) |
| | | | | | | $H_c$ ($H_c$) | $B_r/\mu_o H_c$ ($B_r/H_c$) | | | |
| magnetic field strength H | A/m (Oe) | magnetic flux density B | 400 (5) | 0.5 (5000) | 0.2 (2000) | 20 (0.25) | $10^4$ | $< S6$ | ferrite; soft | Flinn et al (1990) |
| magnetic field strength H | A/m (Oe) | magnetic flux density B | 400 (5) | 0.2 (2000) | 0.2 (2000) | 80 (1) | $10^3$ | $< S6$ | ferrite; soft | Flinn et al (1990) |
| magnetic field strength H | A/m (Oe) | magnetic flux density B | $16 \times 10^4$ (2000) | 0.6 (6000) | 0.4 (6000) | $13 \times 10^4$ (**) (1600) | $10^0$ | $S1$ | permanent-magnet ferrite; hard | Flinn et al (1990) |

(*) This is a slanted curve hence a multiplicative factor used: $6 \times B_r$.
(**) The factor $10^4$ is obviously missing in the scale [amps/m] used for the X-axis in Fig. 21.3(c) of Flinn et al (1990).

In order to plot a 'usable' graph B vs H (M vs H) at least the first quadrant of hysteresis loop indicating the two points $B_r$ ($M_r$) and $H_c$ ($H_{ci}$) must fit into a standard textbook page. Hence the size of the graph is approximately given by the values of $B_r$ ($M_r$) and $H_c$ ($H_{ci}$) and thus different re-scaling factors are required for different materials. An approximate value of the suitable re-scaling factor can be obtained by calculating the ratio $B_r / H_c$ (CGS) or $B_r / \mu_o H_c$ (SI) in the given standard units. This method works well for the narrow and straight hysteresis loops (soft materials) and the wide ones (hard materials) for which the values of $B_r$ ($M_r$) are not too different from their 'saturation' values. For a slanted hysteresis loop, like in Fig. 4 - the B type, with the values of $B_r$ ($M_r$) much smaller than their 'saturation' values a multiplicative factor of between 2 to 6 can be applied to $B_r$ ($M_r$), or alternatively the values of $B_s$ ($M_s$) may be used, if available. Let us illustrate the effect of the re-scaling factors by adopting the unit elements for the y-axis ($y_{unit}$) and x-axis ($x_{unit}$) of equal length, say e.g. one centimeter. Thus if the ratio $B_r / H_c$ (CGS) is, e.g. of the order of (i) $10^3$ or (ii) $10^4$, the suitable re-scaling factor for the graph would be (i) 0.001 or (ii) 0.0001. Such graphs without re-scaling, i.e. using the 1 : 1 unit labeling on the y- and x-axis, would require the maximum on the y-axis, $Y_{max}$, of not less than (i) 10 m - the height of an average four-storey building or (ii) 100 m - one-third of the height of Eiffel Tower in Paris. On such graphs the *actual* inclination of the B-lines after saturation would be exactly 45°. The only drawback would be that they could not be fitted into any textbook. By squeezing the graphs along the y-axis (i) 1000 or (ii) 10000 times, a 'usable' size of the graph is obtained. BUT then the corresponding *apparent* inclination almost vanishes to (i) 0.057° - as for the S6 line in Fig. 3 or (ii) 0.0057° - which cannot be discernibly indicated in Fig. 3. However, the *apparent* nearly zero inclination of the B-lines after saturation, e.g. of the type S5 and S6 in Fig. 3, should not be confused with the exactly zero inclination of the M-lines after saturation independent of the scale used.

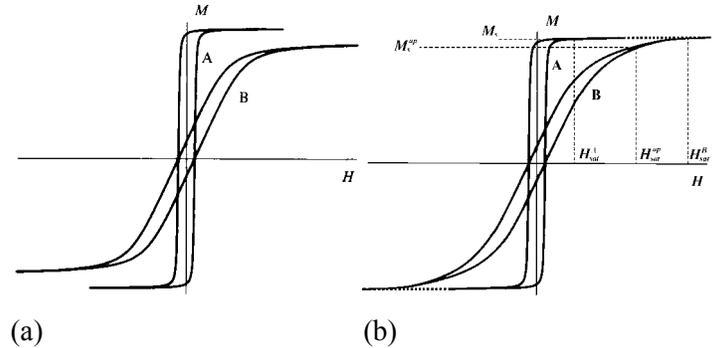

(a) (b)

**Fig. 4.** (a) Two extreme cases of the dependence of magnetization for a soft magnetic material on the angle between the applied field **H** and the easy magnetization direction: the loop (A) - **H** parallel to the easy direction and (B) - **H** perpendicular to the easy direction (adapted from Jakubovics, 1994); (b) Modified loop (B) indicating the apparent saturation $M_{sat}^{ap}$ and the full magnetization saturation at the level of the loop (A).

It is worthwhile to analyse some real data and determine the expected inclination of the B vs H line after the magnetization saturation. This will help (a) clarifying the distinction between the hard and soft magnets in this respect and (b) illustrate the difference between the *actual* and *apparent* inclination. Let us consider the re-scaling factors required to fit the B vs H graphs into a textbook page as described above for three data sets.

*Firstly,* we consider the materials listed in Table 1 (S & R, 2002). To draw a hysteresis graph one may adopt the ratio $S = y_{unit} : x_{unit}$ between 1 and 0.5, since the data points ($H_{ci}$, $H_c$ and $B_r$) are of the same order of magnitude. Hence, the expected inclination of the B-lines after saturation should be **noticeable** - between 45° and 26.6° corresponding to a straight line in-between the lines S1 and S2 in Fig. 3.

*Secondly,* we analyse the textbook B vs H hysteresis diagrams on which the unit elements are indicated. The pertinent data extracted from textbooks are listed in Table 1 together with the values estimated by us. The numerical data for the magnetic materials listed include: (i) the maximum values represented on the x- and y-axis denoted as $X_{max}$ and $Y_{max}$, respectively, (ii) the values of $B_r$ and $H_c$ read out approximately from the graphs, (iii) the ratio $B_r / H_c$, and (iv) the approximate inclination suitable for a given graph. It turns out that in most hysteresis diagrams the inclination of the B-line after saturation should be very small (Table 1). Two exceptions concern Fig. 21.3 (c) in Flinn & Trojan (1990) and Fig. 29.29 (b) in Lea & Burke (1997), where the data for a hard magnetic material should yield a **noticeable** inclination of about $14^o$ and $17^o$, respectively.

*Thirdly,* since almost all data in Table 1 pertain to the soft materials, the data for the hard materials from Table 17.2 in Hummel (1993) are considered. The results collected in Table 2 show convincingly that for the hard materials an appreciable inclination of the B-line after saturation should appear on the B vs H graphs if plotted to fit into a textbook page.

From the above analysis it is evident that for the strong permanent magnets (for references, see S & R, 2002; Hummel, 1993) the B vs H curve no longer 'levels off' after the magnetization saturation. However, in a number of textbooks the B vs H hysteresis loop resembles the "M vs H" type curve with zero inclination (see Section B above). Generally, no mention is made on the dependence of the shape and the inclination the B-lines after saturation on the scale used, which is specific for the soft and hard magnets. In view of the results in Table 1, such presentation may be justifiable in the case of the older books keeping in mind the data available at that time: see, e.g. Tilley (1976), Williams *et al* (1976), Hudson & Nelson (1982), Sears *et al* (1982), Bueche (1986; using the B vs $\mu_o$H graph), Laud (1987). However, in more recent books it is rather out-dated. Serway *et al* (1997) rather inappropriately differentiate between the hard and soft materials by referring in their Fig. 12.5 to a 'wide' hysteresis curve with zero inclination of the B-lines after saturation and to a 'very narrow' hysteresis curve with noticeable inclination, respectively.

Concerning the second version of the category C misconceptions, let us first note that valid cases of a noticeable inclination may appear on the M vs H graphs for strongly anisotropic magnetic materials as discussed briefly in the section D below. For a full discussion of these cases and references, see S & R (2002). One has to keep in mind that this **apparent** 'inclination' applies only to the range between the easy axis saturation and the full saturation (see Fig. 4 (b)). This is distinct from the cases of the schematic M vs H graphs with the shape of the hysteresis loops resembling closely the shape of the B vs H loops and the M-lines after saturation with a noticeable inclination, like in Fig. 1(b). Such inappropriate M vs H graphs occur in a few textbooks: Omar (1975) - most pronounced case, Keer (1993) - the $M_s$ dotted line in Fig. 5.11 is indicated incorrectly with non-zero inclination but the description in text is correct, Halliday *et al* (1992) - slight non-zero inclination while $M_s$ not indicated. These cases cannot be justified by the anisotropic properties of materials since no proper explanations are provided by the authors.

**Table 2.** Characteristics of the hysteresis loop for some permanent magnets; the ratios $B_r$ (remanence) / $H_c$ (coercivity) or the equivalent ones are given as the order of magnitude only; $B_r$ and $H_c$ values are taken from Table 17.2 of Hummel (1993); type of the inclination after saturation refers to the lines in Fig. 3.

| $B_r$ [kG] | $H_c$ [Oe] | $B_r / H_c$ | $B_r$ [T] | $H_c$ [A/m] | $B_r / \mu_o H_c$ | Inclination type | Material name |
|---|---|---|---|---|---|---|---|
| 3.95 | 2400 | 1.6 | 0.4 | $1.9 \times 10^5$ | 1.7 | S1 | Ba-ferrite ( $BaO \cdot 6Fe_2O_3$ ) |
| 6.45 | 4300 | 1.5 | 0.6 | $3.4 \times 10^5$ | 1.4 | S1 | PtCo (77 Pt, 24 Co ) |
| 13 | 14000 | 0.9 | 1.3 | $1.1 \times 10^6$ | 0.9 | S1 | Iron-Neodymium-Boron ( $Fe_{14}Nd_2B_1$ ) |
| 9 | 51 | 176 | 0.9 | $4 \times 10^3$ | 179 | S5 | Steel ( Fe-1%C ) |
| 13.1 | 700 | 19 | 1.3 | $5.6 \times 10^4$ | 18 | S4 ~ S5 | Alnico 5 DG ( 8 Al, 15 Ni, 24 Co, 3 Cu, 50 Fe ) |
| 10 | 450 | 22 | 1 | $3.6 \times 10^4$ | 22 | S4 ~ S5 | Vically 2 ( 13V, 52 Co, 35 Fe) |

### D. Misconceptions concerning the dependence of the shape of the hysteresis loops on the direction of the applied field

Another source of confusion may arise due to the fact that the values of $B_r$ and $H_c$ for the same material may be noticeably different for different physical conditions to which a given material may be subjected. Even the same chemically material may be behave either as a magnetically soft or hard material, depending on the physical conditions applied. These aspects will be discussed in detail in S & R (2002). Here let us consider the possible different shapes of the hysteresis loop as illustrated in Fig. 4 for soft materials. According to Jakubovics (1994) the M vs H loops (A) and (B) in Fig. 4 (a) are for the same material, but the loop (A) corresponds to a greater permeability and saturation value $M_s$ than the loop (B). A curious student encountering in various books one of the two distinct types of the hysteresis loop, i.e. in one book the A-type graph and in another book the B-type graph - each supposedly for the **soft** materials, would certainly be puzzled if no proper explanation on the physical reasons of such difference is provided. This unfortunately is the case in some textbooks (see below). In fact, on either graph: B vs H or M vs H, the loop (A) is obtained if the field **H** is applied **parallel** to the easy direction (ED) of magnetization in the material, whereas the loop (B) is obtained if the field **H** is applied either **parallel** to the hard direction (HD) of magnetization or **perpendicular** to the easy direction (see, e.g., Cullity, 1972; Jakubovics, 1994; den Broeder & Draaisma, 1987; Babkair & Grundy, 1987). One must be careful not to confuse the meaning of 'parallel' and 'perpendicular' directions, since in some materials the easy direction is *perpendicular* to the film surface, and thus the notation used on the graphs: $\perp$ and $\parallel$ - referred to the film surface, may be easily confused with the case: **H** $\parallel$ ED (S & R, 2002).

It appears from our text survey that the hysteresis loops, both the type: B vs H and M vs H, for magnetically soft materials are usually presented in a simplified way as either the loop (A) or (B) in Fig. 4 (a), with the loop (A) being used more often, especially in the less advanced level texts. The confusing point is that the loops (B) appear on the B vs H graphs as an illustration of the distinction between the soft and hard materials represented by a narrow slanted loop (B) and a wide rather straight loop (A), respectively, see, e.g., Knoepfel (2000), John (1983), Tipler (1991), Giancoli (1991, 1989), Whelan & Hodgson (1982), and Akrill *et al* (1982). In view of the possibility of obtaining for soft materials also the narrow straight loops - like the loop (A), having no mention about this possibility arising from the anisotropic properties of the soft materials, such description amounts to a partial truth and may lead to confusion. Thus both the (A) and (B) types of hysteresis loops are physically possible for soft materials, however, the values of the remanence corresponding to each loop are markedly different. Without clearly stating the physical conditions applicable to each loop, a simplified description may not be enough for undergraduate or lower form students to learn properly the properties of the soft ferromagnetic materials. Comparison of the two curves appearing separately in various textbooks may create an incorrect impression concerning the physical situation applicable to each case.

An *additional* misconception arises when the two types of the hysteresis loop (A) and (B) are presented on the same M vs H diagram as in Fig. 4 (a) for the **same** material (see, e.g. Jakubovics, 1994). For the loop (B), at first the **apparent** saturation at a lower value $H_{sat}^{ap}$ is achieved, which correspond to a *full internal* saturation but along the easy magnetization direction. The **full** saturation is achieved only if the field is further increased, which brings about rotation of the magnetization from the easy direction to the direction of the field, which is completed at $H_{sat}^{B}$. This then corresponds to the full saturation of the magnetization at the same level as for the loop (A) as illustrated in the modified Fig. 4 (b). The apparent saturation may correspond to 70 % of the full saturation (see, e.g. Giancoli (1991)) and fields $H_{sat}^{B}$ of up to several times higher than $H_{sat}^{A}$ are required to achieve the full saturation $M_s$. A survey of research papers (S & R, 2002) reveals several examples of the modified hysteresis loop (B) shown in Fig. 4 (b). It turns out that in the experimental practice the full saturation with the field in the hard magnetization direction is rarely achieved. Hence the experimental loops often indicate an **apparent** inclination after $H_{sat}^{ap}$ similar to the B-line inclination on the B vs H graphs discussed in Section C. These two physically distinct cases should not be confused each with other. For a more detailed consideration and discussion of other factors affecting the shape of the hysteresis loop, which is beyond the scope of this paper, see, S & R (2002).

### E. *Misconceptions arising from the hysteresis loops for both soft and hard materials presented in the same figure or using the same scale*

Hysteresis loops for both soft and hard magnetic materials are also found in some texts plotted schematically for comparison either in the same diagram (see, e.g. Murray, 1993) as in Fig. 5 or in two related diagrams using the same scale (see, e.g. Chalmers, 1982 - see also Section F; Budinski, 1996; Budinski & Budinski, 1999). Normally, such presentation, if physically valid, may help to convey better a clear picture to students. However, an opposite effect is achieved, if there exists a great difference in the scale for the two curves plotted in this way, as, e.g. in Fig. 5. If these differences are mentioned neither in the text nor in the figure caption, the diagrams like in Fig. 5 give rather a wrong impression to students. In 1949 Kittel (as quoted by Livingston (1987)) reported that the values of $H_{ci}$ for the hardest and softest ferromagnetic materials differ by a factor of $5 \times 10^6$ (in the SI units). In the recent decades, the range of this difference has grown up to $10^8$ (Livingston, 1987). In fact, comparison of the data in Table 1 and 2 reveals that the values of $H_c$ for the hard and soft magnetic materials listed therein differ by from several hundreds times in the CGS units (or 3 orders of magnitude in the SI units) to 10 thousands times or more (or 6 orders of magnitude in the SI units). However, it appears from figures like Fig. 5 that the coercive force $H_c$ for hard magnetic materials is just only several times larger than that for the soft ones. This kind of misleading comparison appear, e.g. in the textbooks by Arfken *et al* (1984), Brown *et al* (1995), Callister (1994), Geddes (1985), John (1983), Murray (1993), Nelkon & Parker (1978), Ralls *et al* (1976), Schaffer *et al* (1999), Shackelford (1996), Smith (1993 - note that mixed units are indicated in Fig. 15.21(c): SI units for the y-axis, whereas CGS units for the x-axis), Turton (2000), Whelan & Hodgson (1982).

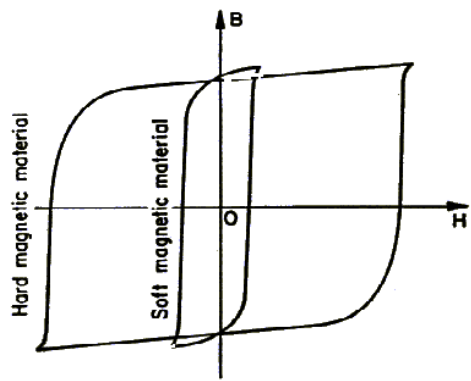

**Fig. 5.** The hysteresis loop for soft and hard magnetic materials plotted in the same graph for comparison (adapted from Murray, 1993).

*F. Other problems concerning terminology*

Various minor problems concerning confusing terminology have also been identified in our survey. Table 1 indicates as wide variety of naming conventions and symbols used. A greater uniformity in this respect, i.e. adherence to the standard nomenclature and units, would help avoiding confusion. Students may easily confuse the various terms used for the applied field H and magnetic induction B. As Table 1 indicates other names used are, e.g., for H: current, magnetic intensity, magnetic field strength, magnetizing field, while for B: magnetic field, flux density, total or internal magnetic field. Chalmers (1982) uses the intensity of magnetization *I*, which is defined only in the SI (Kennelly) unit system (Jiles, 1991). We have carried out a quick survey among several physics MPhil students at CityU asking them to identify the term (i.e. magnetic field, magnetic field, and magnetic field strength) corresponding to H and B. As we expected, they can distinguish between the physical meaning of H and B, but they rather mix up the corresponding names. To a certain extent, this finding is a reflection of the unhealthy situation prevailing in the textbooks (see Table 1). The existing variety of conventions used for the quantities **H**, **B** and **M** hampers the understanding of physics, especially even more seriously for the lower form students.

The improper use of the term *'polarization'* to interpret the B vs H curve occurs in Abele (1993). Normally, *'polarization'* is used to describe the electric quantities rather than the magnetic ones. *'Polarization'* of dielectric materials is an analogue quantity to *'magnetization'* in magnetism, however, the two terms are not equivalent. Besides, John (1983) denotes on the y-axis the magnetization confusingly as B, *"magnetization induced B"*, and describes the x-axis as: *"magnetizing force H"*. Improper use of symbols leads to confusion like, e.g., *"saturation magnetization"* denoted as *"$B_s$"* by Murray (1993). Another confusing notion is used by Granet (1980): *"magnetization current"*, which means the electric current, which induces H, which, in turn, magnetizes the sample. Such terminology mixing up the magnetic and electrical notions and/or quantities may be confusing for students and should be avoided for pedagogical reasons. Another example of confusion is: *"If the field intensity is increased to its maximum in this direction, then reversed again ..."* (Thornton & Colangelo, 1985). In fact, there is no 'maximum limit' for the applied field, apart from the limits imposed by the experimental equipment.

## 4. A survey of students' understanding of the hysteresis loop

The effect of the confusion and misconceptions existing in textbooks on the students' understanding of the hysteresis loop can be assessed by analysing the results of examinations or tests. In our lecture notes for the condensed matter physics course (Rudowicz, 2000; available from CZR upon request) we have presented clearly, so briefly, the distinction between the pertinent notions as well as warned students about the misinterpretations discussed in Section A above. However, the analysis of the results of the CMP examination carried out in May 2001 indicates that the message has not reached some students. Eleven students out of total 15 attempted the two questions concerning the hysteresis loop, stated as follows:

a) *Draw a schematic diagram of the magnetic field intensity inside a material **B** versus the (external) applied magnetic field strength **H** for an initially unmagnetized ferromagnetic material.*

b) *Define the following quantities: (1) the saturation magnetic field, (2) the remanence, and (3) the coercive force. Indicate these quantities on the diagram **B** versus **H** (the hysteresis loop).*

Most of them performed not too well since they often misinterpreted the characteristics of the hysteresis loop. Common mistakes include, e.g., (a) mixing up the coercivity with the remanence, (b) indicating a 'maximum' value of the applied field H and the magnetic induction B, (c) stating that both B and M are equal to zero at $H_c$. Obviously, some of these misconceptions concerning the hysteresis loop are quite close to the ones existing in the surveyed textbooks as discussed above. However, such misconceptions by our students may not originate from any insufficient clarifications of the major terms in the books they might have used, but rather are due to the students' attitudes to learning. In general, a good interpretation of a particular topic in textbooks may help teachers to increase the efficiency of teaching (and save their time which would have been used for clarifications), whereas students to improve their understanding of the physical concepts beyond the level presented at the lectures. On the contrary, the improper definitions of the crucial terms and/or outright misconceptions will most certainly hamper the students' understanding and may contribute to a reduced interest in further physics studies, especially at a lower level of students' education (Hubisz, 2000).

## 5. Conclusions and suggestions

It appears that the two possible ways of presenting the hysteresis loop for ferromagnetic materials, **B vs H** and **M vs H**, are, to a certain extent, confused each with other in several textbooks. This leads to various misconceptions concerning the meaning of the physical quantities as well as the characteristic features of the hysteresis loop for the soft and hard magnetic materials. We suggest that the name *'coercive force'* (or *'coercivity'*) and the symbol $H_c$, correctly defined for the B vs H curve, should not be used if referred to the M vs H curve. Using in the latter case the adjective *'intrinsic'* and the symbol $H_{ci}$ is strongly recommended. It may help avoiding the misconceptions discussed above and reduce the present confusion widely spread in the textbooks. Hence the authors and editors should pay more attention to proper definitions of the terms involved. Interestingly, among the books by Beiser (1986, 1991, 1992), the book (1986) belongs to the *'misinterpretation sample'*, while the two later books (1991, 1992) are correct in this aspect. It is hoped that by bringing the problems in questions to the attention of physics teachers and students, the correct interpretation of the hysteresis loop will prevail in future.

Our survey of textbooks reveal several deeper pedagogical issues related to the presentation of the hysteresis loop, which may apply to various other topics as well. One is the distinction between the 'exact' and 'approximate' quantities and the related description of a physical situation. In the present case we have considered the approximation 'H small as compared with M', leading to $B \approx \mu_o M$ and $H_c \approx H_{ci}$ for soft magnetic materials. If the conditions for which a given approximation is valid are not clearly stated, the approximate picture may be implicitly taken as a representation of the exact situation. The consequences of such misleading approach may be wide-ranging - from imprinting misconceptions, i.e. false images, in the students' minds to misinterpretation of the properties of one class of materials (here, soft magnets) as being equivalent to those of another class (here, hard magnets).

The inherent danger in using *'schematic'* diagrams for presentation of the dependencies between various physical quantities is another important issue. Having no units and values provided for the y- and x-axis constitutes a detachment from a real physical situation. It may not only hamper students' understanding of the underlying physics, but also lead to false impressions about the relationships between the quantities involved and, in consequence, create misconceptions. This is best exemplified for the present topic by Fig. 4 and Fig. 5. The drawbacks of 'schematic' representation of each hysteresis loops are compounded by the 'space saving' and using a combined diagram like in Fig. 5, which implies the same limits and values are applicable for both types of magnetic materials. As we amply illustrated above this is far from the true situation. Schematic diagrams which do not reflect correctly the underlying physical situation become a piece of graphic art only. Providing neither symbols nor description of the quantities on the x- and y-axis of a graph (see, e.g. Fig. 15.9 in Machlup, 1988) should also be avoided in physics text as an inappropriate from both scientific and pedagogical point of view.

Finally, let us mention the idea of creating a website listing errors and misconceptions in textbooks. The individual lecturers could add up their knowledge in this respect to a well organized structure listing various topics. We have had preliminary talks within CityU about the idea of having such a website residing on the CityU computer network but the response was rather muted. Our initial Internet search for the keywords: *'errors', 'misprints', 'corrigenda', 'errata',* has, however, revealed, no relevant sites. A similar idea was proposed by Hubisz (2000) concerning science textbooks. Interestingly we have located this website due to letter in American Physical Society Newsletter (April, 2001, p.4). Since the URL address was misprinted, we have tracked this site down via the university name (North Carolina State University). Only recently by chance we have learnt of the existing website listing errors in physics textbooks:

(http://www.escape.ca/~dcc/phys/errors.html). It appears that the benefits of such website for teachers and students in improving general understanding of physics may be substantial.

## Acknowledgements

This work was supported by the City University of Hong Kong through the research grant: QEF # 8710126.

## References

Abele M G 1993 *Structures of Permanent Magnets Generation of Uniform Fields* (New York: John Wiley & Sons) p 33-35

Aharoni A 1996 *Introduction to the Theory of Ferromagnetism* (Oxford: Clarendon Press) p 1-3

Akrill T B, Bennet G A G and Millar C J 1982 *Physics* (London: Edward Arnold) p 234

Anderson H L 1989 *A Physicist's Desk Reference Physics Vade Mecum* (New York: American Institute of Physics)

Anderson J C, Leaver K D, Rawlings R D and Alexander J M 1990 *Materials Science* (London: Chapman and Hall) p 501-502

Anderson J P and Blotzer R J 1999 *Permeability and hysteresis measurement* The Measurement, Instrumentation, and Sensors Handbook ed J G Webster (Florida: CRC Press) p 49-6-7

Arfken G B, Griffing D F, Kelly D C and Priest J 1984 *University Physics* (Orlando: Academic Press) p 694-697

Arrott A S 1983 *Ferromagnetism* in Concise Encyclopedia of Solid State Physics ed R G Lerner & G L Trigg (London: Addison-Wesley Publishing) p 97-101

Babkair S S and Grundy P J 1987 *Multilayer ferromagnetic thin film supersturctures -Co/Cr* in Proceedings of the International Symposium on Physics of Magnetic Materials ed M Takahashi et al (Singapore: World Scientific) p 267-270

Barger V D and Olsson M G 1987 *Classical Electricity and Magnetism* (Boston: Allyn and Bacon) p 318-319

Beiser A 1986 *Schaum's outline of Theory and Problems of Applied Physics* (Singapore: McGraw-Hill ) p 195-197
Beiser A 1991 *Physics* (Massachusetts: Addison-Wesley) p 546-548
Beiser A 1992 *Modern Technical Physics* (Massachusetts: Addison-Wesley) p 576-578
Besancon R M 1985 *The Encyclopedia of Physics* (New York: Van Nostrand Reinhold) p 440
Brick R M, Pense A W and Gordon R B 1977 *Structure and Properties of Engineering Materials* (New York: McGraw-Hill) p 33-34
Brown W, Emery T, Gregory M, Hackett R and Yates C 1995 *Advanced Physics* (Singapore: Longman) p 226-227
Buckwalter G L and Riban D M 1987 *College Physics* (New York: McGraw-Hill Book Company) p 511-512
Budinski K G 1996 *Engineering Materials Properties and Selection* (New Jersey: Prentice Hall) p 28
Budinski K G and Budinski M K 1999 *Engineering Materials Properties and Selection* (New Jersey: Prentice Hall) p 29-30
Bueche F J 1986 *Introduction to Physics for Scientists and Engineers* (New York: Glencoe/McGraw- Hill) p 522
Burke H E 1986 *Handbook of Magnetic Phenomena* (New York: Van Nostrand Reinhold) p 63-64
Callister, Jr W D 1994 *Material Science and Engineering An Introduction* (New York: John Wiley & Sons) p 673-675
Chalmers B 1982 *The Structure and Properties of Solids* (London: Heyden) p 46-49
Compton A J 1986 *Basic Electromagnetism and its Applications* (Berkshire: Van Nostrand Reinhold) p 97-98
Cullity B D 1972 *Introducton to Magnetic Materials* (Massachusetts: Addison-Wesley) p 18-19
Daintith J 1981 *Dictionary of Physics (New York: Facts On File)* p. 88
Dalven R 1990 *Introduction to Applied Solid State Physics* (New York: Plenum Press) p 367-376
den Broeder F J A and Draaisma H J G 1987 *Structure and magnetism of polycrystalline multilayers containing ultrathin Co or Fe* in Proceedings of the International Symposium on Physics of Magnetic Materials ed M Takahashi et al (Singapore: World Scientific) p 234-239
Donoho P L 1983 *Hysteresis* in Concise Encyclopedia of Solid State Physics ed R G Lerner & G L Trigg (London: Addison-Wesley Publishing) p 122-123
Dugdale D 1993 *Essential of electromagnetism* (New York: American Institute of Physics) p 198-199
Elliott R J and Gibson A F 1978 *An Introduction to Solid State Physics and its Applications* (London: English Language Book Society) p 464-466
Elliott S R 1998 *The Physics and Chemistry of Solids* (Chichester: John Wiley & Sons) p 630
Elwell D and Pointon A J 1979 *Physics for Engineers and Scientists* (Chichester: Ellis Horwood) p 307-308
Fishbane P M, Gasiorowicz S and Thornton S T 1993 *Physics for Scientists and Engineers* (New Jersey: Prentice Hall) p 947-948.
Flinn R A and Trojan P K 1990 *Engineering Materials and Their applications* (Dallas: Houghton Mifflin) p S178-S182
Geddes S M 1985 *Advanced Physics* (Houndmills: Macmillan Education) p 57-59
Giancoli D C 1989 *Physics for Scientists and Engineers with Modern Physics* (New Jersey: Prentice Hall) p 662-663
Giancoli D C 1991 *Physics Principles with Applications* (New Jersey: Prentice Hall) p 529-531
Granet I 1980 *Modern Materials Science (Virginia: Prentice-Hall)* p 422-427
Grant I S and Phillips W R 1990 *Electromagnetism* (Chichester: John Wiley & Sons) p 201-242
Gray H J and Isaacs A 1975 *A New Dictionary of Physics (London: Longman)* p 268
Halliday D, Resnick R and Krane K S 1992 *Physics* Vol.2 (New York: John Wiley & Sons) p 814
Hammond P 1986 *Electromagnetism for Engineers An Introductory Course* (Oxford: Pergamon press) p 134-137
Harris N C and Hemmerling E M 1980 *Introductory Applied Physics* (New York: McGraw-Hill) p 570-571
Hickey & Schibeci 1999 *Phys. Educ.* **34** 383-388
Hoon S R and Tanner B K 1985 *Phys. Educ.* **20** 61-65
Hubisz J L 2000 *Review of Middle School Physical Science Texts* (http://www.psrc-online.org/curriculum/book.html) - accessed June 2001 p 1-98
Hudson A and Nelson R 1982 *University Physics* (San Diego: Harcourt Brace Jovanovich) p 669

Hummel R E 1993 *Electronic Properties of Materials* (Berlin: Springer-Verlag) p 314, 319
Jakubovics J P 1994 *Magnetism and Magnetic Materials* (Cambridge: The Institute of Materials)
Jastrzebski Z D 1987 *The Nature and Properties of Engineering materials* (New York: John Wiley & Sons) p 482-485
Jiles D 1991 *Introduction to Magnetism and Magnetic Materials* (London: Chapman & Hall) p 70-73
John V B 1983 *Introduction to Engineering Materials* (London: Macmillan) p 130-131
Keer H V 1993 *Principles of the Solid State* (New York: John Wiley and Sons) p 235-236
Kittel C 1996 *Introduction to Solid State Physics* (New York: John Wiley & Sons) p 468-470
Knoepfel H E 2000 *Magnetic fields A Comprehensive Theoretical Treatise for Practical Use* (New York: John Wiley & Sons) p 486-491
Lapedes D N 1978 *Dictionary of Physics and Mathematics* (New York: McGraw-Hill) p 469
Laud B B 1987 *Electromagnetics* (New York: John Wiley & Sons) p 162-163
Lea S M and Burke J R 1997 *Physics The Nature of Things* (New York: West Publishing) p 941-942
Lerner R G and Trigg G L 1991 *Encyclopedia of Physics* (New York: VCH Publishers) p 692-693
Levy R A 1968 *Principles of Solid State Physics* (New York: Academic Press) p 258-260
Livingston 1987 *Upper and lower limits of hard and soft magnetic properties* in Proceedings of the International Symposium on Physics of Magnetic Materials ed M Takahashi et al (Singapore: World Scientific) p 3-16
Lord M P 1986 *Dictionary of Physics* (London: Macmillan) p 140
Lorrain P and Corson D R 1979 *Electromagnetism* (San Francisco: W.H. Freeman and Company) p 342-345
Lovell M C, Avery A J and Vernon M W 1981 *Physical Properties of Materials* (New York: Van Nostrand Reinhold) p 189
Machlup S 1988 *Physics* (New York: John Wiley & Sons) p 459
Meyers R A 1990 *Encyclopedia of Modern Physics* (San Diego: Harcourt Brace Jovanovich) p 254-255
Murray G T 1993 *Introduction to Engineering Materials Behavior, Properties, and Selection* (New York: Marel Dekker) p 529-531
Nelkon M and Parker P 1978 *Advanced Level Physics* (London: Heinemann Educational Books) p 843
Omar M A 1975 *Elementary Solid State Physics: Principles and Applications* (Massachusetts: Addison-Wesley) p 461
Ouseph P J 1986 *Technical Physics (New York: Delmar)* p 537-538
Parker S P 1993 *Encyclopedia of Physics* (New York: McGraw-Hill) p 733
Pitt V H 1986 *The Penguin Dictionary of Physics* (Middlesex: Penguin books) p 186-187
Pollock D D 1985 *Physical of Materials for Engineers* Vol. II (Boca Raton: CRC Press) p 138-139
Pollock D D 1990 *Physics of Engineering Materials* (New Jersey: Prentice Hall) p 587-589
Ralls K M, Courtney T H and Wulff 1976 *Introduction to Materials Science and Engineering* (New York: John Wiley & Sons) p 575-577
Rhyne J J 1983 *Magnetic Materials* in Concise Encyclopedia of Solid State Physics ed R G Lerner & G L Trigg (London: Addison-Wesley Publishing) p 160-162
Rogalski M S and Palmer S B 2000 *Solid-State Physics* (Australia: Gordon and Breach Science Publishers) p 379
Rudowicz C, 2001, *Lecture Notes: Condensed Matter Physics*, City University of Hong Kong, unpublished.
Schaffer J P, Saxena a, Antolovich S D, Sanders Jr. T H and Warner S B 1999 *The Science and Design of Engineering Materials* (Boston: McGraw-Hill) p 527-530
Sears F W, Zemansky M W and Young H D 1982 *University Physics* (California: Addison-Wesley) p 673-674
Selleck E 1991 *Technical Physics* (New York: Delmar) p 879-880
Serway R A 1990 *Physics for Scientists & Engineers with Modern Physics* (Philadelphia: Saunders College) p 857-859
Serway R A, Moses C J and Moyer C A 1997 *Modern Physics* (Fort: Saunders College) p 481-483
Shackelford J F 1996 *Introduction to Materials Science for Engineers* (New Jersey: Prentice Hall) p 507-512
Skomski R and Coey J M D 1999 *Studies in Condensed Matter Physics Permanent Magnetism* (Bristol: Institute of Physics) p 169-174
Smith W F 1993 *Foundations of Materials Science and Engineering* (New York: McGraw-Hill) p 827-828
Sung H W F and Rudowicz C 2002 J. Mag. Magn. Mat. - submitted Feb 2002

Thornton P A and Colangelo V J 1985 *Fundamentals of Engineering Materials* (New Jersey: Prentice-Hall) p 372-373
Tilley D E 1976 *University Physics for Science and Engineering* (California: Cummings Publishing) p 532-534
Tipler P A 1991 *Physics for Scientists and Engineers* (New York: Worth) p 886-888
Turton R 2000 *The Physics of Solids* (Oxford: Oxford University Press) p 237
Van Vlack L H 1970 *Materials Science for Engineers* (Menlo Park: Addison-Wesley) p 327-328
Van Vlack L H 1982 *Materials for Engineering: Concepts and Applications* (Menlo Park: Addison-Wesley) p 544
Vermariën H, McConnell E and Li Y F 1999 *Reading / recording devices* The Measurement, Instrumentation, and Sensors Handbook ed J G Webster (Florida: CRC Press) p 96-24-25
Wert C A and Thomson R M 1970 *Physics of Solids* (New York: McGraw-Hill) p 455-456
Whelan P M and Hodgson M J 1982 *Essential Principles of Physics* (London: John Murray) p 429
Williams J E Trinklein F E and Metcalfe H C 1976 *Modern Physics* (New York: Holt, Rinehart and Winston) p 468
Wilson I 1979 *Engineering Solids* (London: McGraw Hill) p 119-123

**Appendix I.  List of other surveyed textbooks not included in the references**

The list of textbooks surveyed, in which no relevant misconceptions and/or confusions were identified and which are not quoted in text, is given below.

Benson H 1991 *University Physics* (New York: John Wiley & Sons) p 653
Blakemore J S 1985 *Solid State Physics* (Cambridge: Cambridge University Press) p 450
Bube R H 1992 *Electrons in Solids An Introductory Survey* (Boston: Academic Press) p 261-262
Burns G 1985 *Solid State Physics* (Orlando: Academic Press) p 614
Christman J R 1988 *Fundamentals of Solid State Physics* (New York: John Wiley & Sons) p 369
Coren R L 1989 Basic Engineering Electromagnetics An Applied Approach (New Jersey: Prentice Hall) p 76-77
Craik D 1995 *Magnetism Principles and Applications* p 105
Crangle J 1991 *Solid State Magnetism* (London: Edward Arnold) p 168-171
Dekker A J 1960 *Solid State Physics* (London: Macmillan & Co) p 476
Enz C P 1992 *Lecture Notes in Physics vol. 11 A Course on Many-Body Theory Applied to Solid-State Physics* (Singopore: World Scientific) p 267
Feynman Leighton and Sands 1989 *Commemorative Issue The Feynman Lectures on Physics Vol. II* (Addison-Wesley: California) p 36-5-36-9
Gershenfeld N 2000 *The Physics of Information Technology* (Cambridge: Cambridge University Press) p 193
Guinier A and Jullien R 1989 *The Solid State From Superconductors to Superalloys* (Oxford: Oxford University Press) p 163-166
Halliday D, and Resnick R 1978 *Physics Part I and II* (New York: John Wiley & Sons) p 827
Hammond P and Sykulski J K 1994 *Engineering Electromagnetism Physical Processes and Computation* (Oxford: Oxford University Press) p 223-225
Hook J R and Hall H E 1991 *Solid State Physics* (Chichester: John Wiley & Sons) p 251
Joseph A, Pomeranz K, Prince J and Sacher D 1978 *Physics for Engineering Technology* (New York: John Wiley & Sons) p 442
Kahn O 1999 *Magnetic anisotropy in molecule-based magnets* in Metal-Organic and Organic Molecular Magnets ed P Day and A E Underhill (Cambridge: Royal Society of Chemistry) p 150-168
Kinoshita M 1999 *Molecular-based magnets: setting the scene in* Metal-Organic and Organic Molecular Magnets ed P Day and A E Underhill (Cambridge: Royal Society of Chemistry) p 4-21
Lorrain P and Corson D R 1979 *Electromagnetism* (San Francisco: W.H. Freeman and Company) p 342-345
Marion J B and Hornyak W F 1984 *Principles of Physics* (Philadelphia: Saunders College Publishing) p 750
Myers H P 1991 *Introductory Solid State Physics* (London: Taylor & Francis) p 377
Narang B S 1983 Material Science and Processes (Delhi: CBS) p 66-67
Ohanian H C 1989 *Physics Vol.2* (New York: W.W. Norton) p 818
Parker S P 1988 *Solid-State Physics Source Book* (New York: Mcgraw-Hill) p 223-230
Radin S H and Folk R T 1982 *Physics for Scientists and Engineers* (New Jersey: Prentice-Hall) p 657
Rosenberg H M 1983 *The Sold State* (Oxford: Clarendon) p 202
Rudden M N and Wilson J 1993 *Elements of Solid State Physics* (Chichester: John Wiley & Sons) p 103, 109
Schneider, Jr S J, Davis J R, Davidson G M, Lampman S R, Woods M S and Zorc T B 1991 *Engineered Materials Handbook Vol. 4 (USA: ASM International)* p 1162
Swartz C E 1981 *Phenomenal Physics* (New York: John Wiley & Sons) p 609
Tanner B K 1995 *Introduction to the Physics Electrons in Solids* (Cambridge: Cambridge University Press) p 181-182
Tippens P E 1991 *Physics* (New York: Glencoe/McGraw-Hill) p 671-672
Van Vlack L H 1989 *Elements of Materials Science and Engineering* (Menlo park: Addison-Wesley) p 450-453
Vonsovskii S V 1974 *Magnetism* Vol. One (New York: John Wiley & Sons) p 42-43
Young H D and Freeman R A 1996 *Extended Version with Modern Physics: University Physics* (Massachusetts: Addison-wesley) p 926
Zafiratos C D 1985 *Physics* (New York: John Wiley & Sons) p 673-674